\documentstyle[preprint,aps]{revtex}

\begin{document}
\draft
\author{S. Candia and L. Civale}
\address{Comisi\'on Nacional de Energ\'{\i}a At\'omica, Centro At\'omico
Bariloche and Instituto Balseiro \\
8400 Bariloche, RN, Argentina}
\title{Angular dependence of the magnetization of isotropic superconductors:
which is the vortex direction?}
\maketitle

\begin{abstract}	
We present studies of the dc magnetization of thin platelike samples of the
isotropic type II superconductor PbTl(10\%), as a function of the angle
between the normal to the sample and the applied magnetic field ${\bf H}$.
We determine the magnetization vector ${\bf M}$ by measuring the components
both parallel and normal to ${\bf H}$ in a SQUID magnetometer, and we
further decompose it in its reversible and irreversible contributions.
The behavior of the reversible magnetization is
well understood in terms of minimization of the free energy taking into
account geometrical effects. In the mixed state at low fields, the dominant
effect is the line energy gained by shortening the vortices, thus the flux
lines are almost normal to the sample surface. Due to the geometrical
constrain, the irreversible magnetization ${\bf M}_{irr}$ remains locked to
the sample normal over a wide range of fields and orientations, as already
known. We show that in order to undestand the angle and field dependence of
the modulus of ${\bf M}_{irr}$, which is a measure of the vortex pinning,
and to correctly extract the field dependent critical current density,
the knowledge of the modulus and orientation of the induction field
${\bf B}$ is required.

\end{abstract}

\section{Introduction}

It is well known that sample shape effects have a strong influence in the
magnetic response of superconductors. The discovery of the high temperature
superconductors (HTSC) revived the interest in these effects, particularly
for the case of platelike geometries
\cite{clem-1994,brandt-1994b,schuster93a,frankel-1979,kunchur-et-al-1991,hellman-et-al-1992,zhukov-et-al-1997}
because those materials are frequently produced in the
form of thin films or platelike single crystals.

When we consider the shape effects on the dc magnetization of samples of
type II superconductors whose dimensions are large as compared to the
superconducting penetration length $\lambda $, it is useful to divide the
problem in two parts, namely the reversible magnetization and the
irreversible magnetization that arises from vortex pinning. The first case
appears as a rather simple textbook problem (at least for geometries that
can be approximated by an ellipsoid) that can be solved by using the
demagnetizing tensor formalism \cite{parks,landau-v8}. The second aspect is
more complex, and have received considerable recent attention, both
theoretically and experimentally
\cite{clem-1994,brandt-1994b,schuster93a,frankel-1979,kunchur-et-al-1991,hellman-et-al-1992,zhukov-et-al-1997}.

The macroscopic magnetic response due to flux pinning is well described by
the critical state model \cite{campbell-1972,bean-1962}, according to
which the current density flowing in any part of the sample is either the
critical current density $J_c$ or zero. This current produces an spatially
inhomogeneous magnetic induction ${\bf B}\left( {\bf r}\right) $, that is
detected in magnetometry studies as the non-equilibrium part of the
magnetization. In general, $J_c$ is a function of the temperature $T$ and
the local ${\bf B}\left( {\bf r}\right) $. The simplest critical state
model, proposed by Bean\cite{bean-1962}, assumes that 
$J_c$ is independent of ${\bf B}$, thus resulting in a uniform current 
density when the applied magnetic field ${\bf H}$ is varied in a typical 
isothermal measurement ${\bf M}({\bf H})$ (hysteresis loop).

Even within the simple Bean model, the calculation of
${\bf B}\left({\bf r}\right)$ in thin samples with ${\bf H}$ normal to the sample surface is
rather complex, and has been the subject of extensive modeling and
numerical analysis\cite{clem-1994,brandt-1994b,daeumling-1989}.
The situation is further complicated when the realistic cases of field
dependent and/or anisotropic $J_c$ (e.g., in HTSC) are considered\cite{conner-1991}. 
Several experimental techniques have also been used to measure the local 
${\bf B}$\cite{tamegai-1992,schuster93a}

Another aspect of the problem, that has been less explored, is the magnetic
response when ${\bf H}$ is tilted with respect to the normal to the surface
of thin samples. As the non-equilibrium screening currents are strongly
constrained by the geometry to flow parallel to the sample surface, there is
a large angular range of applied fields in which the irreversible
magnetization ${\bf M}_{irr}$ points almost perpendicular to the
surface \cite{hellman-et-al-1992,zhukov-et-al-1997}, and
consequently ${\bf B}$ is not parallel to ${\bf H}$. This {\it geometrical
anisotropy} results in some angular effects that are qualitatively similar
to those observed in extremely anisotropic (quasi two-dimensional)
superconductors \cite{blatter-et-al-1992,blatter-et-al-1994,buzdin-et-al-1990}.
The understanding of angular effects due to sample
geometry are thus necessary for a proper interpretation of those studies.

The above considerations indicate that it makes sense to explore the angular
behavior of thin samples of {\it isotropic} superconductors. This is the
approach followed, for instance, by Hellman et al. \cite{hellman-et-al-1992}, and more recently by Zhukov et al. \cite{zhukov-et-al-1997}. Those studies
are mostly focused on the angular dependence of the irreversible
magnetization, and little attention is paid to the modulus and angle of the
magnetic induction ${\bf B}$, that represents the actual density and
direction of the vortices.

In this work we show that in order to understand the angle and field
dependence of the irreversible magnetization of thin samples, and to
correctly extract the field dependent $J_c$ using the critical state model,
a complete knowledge
of the vector ${\bf B}$ is required. We present studies of the dc
magnetization of a platelike isotropic (conventional) superconductor as a
function of angle with respect to ${\bf H}$. We measure the total
magnetization vector ${\bf M}$ and decompose it in its reversible and
irreversible parts. We demonstrate that they have different directions, and
analyze the angular and field dependence of both of them. We show that the
geometrical effects on the reversible response determine the vector
${\bf B}$, which is the key variable in the problem.

\section{Experimental details and procedures}

In the present study we used samples of the isotropic superconducting alloy $%
Pb_{0.9}Tl_{0.1}$. Two samples, a circle of $3.3mm$ in diameter and $43\mu m$
thick and a square of $2.4\times 2.4mm^2$ and $39\mu m$ thick were cut from
the same foil. The thicknesses were determined from the samples' weight and
known density ($\sim 11.4g/cm^3$), and later confirmed from the magnetic
measurements. Samples dimensions were chosen so that the aspect ratios are
similar to that of the $YBa_2Cu_3O_{7-\delta }$ crystals that we use. The
critical temperature is $T_c\sim 7.3K$. As results from both samples are
very similar, only data from the square sample will be presented here.

Dc magnetization measurements were made in a Quantum Design magnetometer
equipped with two SQUID detectors, each one coupled to a set of pick up
coils. This allows us to record two perpendicular components of the magnetic
moment, namely the longitudinal component $m_l$ (parallel to ${\bf H)}$, and
the transverse component $m_t$ (normal to ${\bf H)}$, and thus to determine
the total magnetic moment vector ${\bf m}$. Samples can be rotated in situ
around an axis perpendicular to ${\bf H}$ using a home-made sample rotator
device \cite{casa}. The angular resolution, given by the minimum step, is $%
\sim 0.5^{\circ }$, but the overall error in the absolute value of the angle
is about $\sim 2^{\circ }$. The square sample was mounted with one pair of
its sides parallel to the rotation axis.

The initial alignment was performed as follows\cite{casa}. Sample was first
cooled in
zero field below $T_c$ and then a small field was applied. Sample was then
rotated using the home-made system until the angle formed by its normal and $%
{\bf H}$ was about $45^{\circ }$. Subsequently, the sample was rotated
around the vertical axis (parallel to ${\bf H}$) and the transverse
component $m_t$ was measured as a function of that orientation. Once the
maximum in $m_t$ was found, indicating that ${\bf m}$ lies in the plane
formed by ${\bf H}$ and the axis of the transverse coils\cite{qd}, the
angular position around the vertical axis was maintained fixed throughout
the rest of the experiment. (The capability to rotate around the vertical
axis is provided by the commercial system and is not affected by
our perpendicular rotation device).

It is known that the measurement of $m_t$ posses some difficulties which
originate in the appearance of an spurious signal due to the longitudinal
component $m_l$, that is detected by the transverse pick-up coils. This
occurs when the sample is slightly off-centered with respect to the vertical
axis of the coils\cite{qd}, which is frequently the case. We have completely
and satisfactorily solved this problem. The solution includes the external
processing of the original SQUID output signal using a software developed
ad-hoc. All the details related with the hardware and software of the sample
rotation system are presented elsewhere \cite{casa}.

We performed measurements of both components of the dc magnetization, $%
M_l=m_l/V$ and $M_t=m_t/V$ (where $V$ is the sample volume), as a function
of applied field, at fixed temperature and for fixed values of $\Theta _H$,
the angle formed by ${\bf H}$ and the normal to the sample surface, ${\bf n}$
(see sketch in Figure \ref{mlmt}). We repeated these isothermal hysteresis
loops for different $\Theta _H$. After each loop was finished, the sample
was rotated to the new $\Theta _H$, warmed up above $T_c$ and cooled down in
zero field (ZFC)\ before starting the new run. In this way, the initial
Meissner response was recorded for each angle.

\section{Results and discussion}

\subsection{Meissner response as a function of angle}

\label{sec.rota}Figure \ref{mlmt} shows a typical pair of curves of $M_t$
and $M_l$ versus $H$ for the square sample, recorded at $T=4.5K$ using the
procedure described in the previous section. In this particular case $\Theta
_H\simeq 60^{\circ }$. Both $M_t$ and $M_l$ exhibit an initial linear
dependence on $H$ characteristic of the Meissner response, followed by an
hysteretic behavior indicative of vortex pinning when $H$ is further
increased and then decreased.

We first analyze the Meissner response using the standard demagnetizing
tensor formalism \cite{parks,landau-v8}. Let's consider a homogeneous magnetic
material immersed in a uniform field ${\bf H}$. If the shape of the sample
can be approximated by an ellipsoid, then the internal field ${\bf H}_i$, the
magnetization ${\bf M}$ and the magnetic induction ${\bf B=H}%
_i+4\pi {\bf M}$ are all uniform, and the internal field is ${\bf H}_i={\bf %
H+}4\pi {\bf \overline{\overline{\nu }}M}$, where ${\bf \overline{\overline{%
\nu }}}$ is the demagnetizing tensor. We thus have

\begin{equation}
\label{bhm}{\bf B}={\bf H}+4\pi ({\bf \overline{\overline{1}}-\overline{%
\overline{\nu }}}){\bf M} 
\end{equation}
(Note that Eq. \ref{bhm} is valid only if ${\bf M}$ is uniform. It cannot be
used, for instance, to describe the critical state of a superconductor).

We apply Eq. \ref{bhm} to the Meissner state, where ${\bf B}=0$. We
decompose ${\bf H}$ in the two directions ${\bf n}$ and ${\bf p}$, where the
unit vector ${\bf p}$ is defined by the intersection of the sample surface
and the plane formed by ${\bf n}$ and ${\bf H}$ (see sketch in Figure \ref
{mlmt}). Then $\overline{\overline{\nu }}$ becomes diagonal\cite{blatter-et-al-1994,landau-v8}
and we have
$H_{\parallel }=H\sin {\Theta _H}=-4\pi (1-\nu _{\parallel })M_{\parallel }$
and $H_{\perp }=H\cos {\Theta _H}=-4\pi (1-\nu _{\perp })M_{\perp }$. For
very thin samples $\nu _{\parallel }\approx \nu $ and $\nu _{\perp }\approx
1-2\nu $ with $\nu \ll 1$, thus

\begin{eqnarray}
4\pi m_l &=&-\left( \frac 1{2\nu }\cos ^2\Theta _H+\frac 1{1-\nu }\sin
^2\Theta _H\right) V H  \nonumber \\
4\pi m_t &=&-\left( \frac 1{2\nu }-\frac 1{1-\nu }\right) \sin {\Theta _H}%
\cos {\Theta _H}V H \label{ml.mt}\\
&&  \nonumber
\end{eqnarray}

where we have written eqs. \ref{ml.mt} in terms of the measured quantities $%
m_l$ and $m_t$. A simple way to test this dependence is to make a ZFC, then
apply a small field (typically $\sim 10Oe$) and rotate the sample at fixed $%
H $. One of such measurements is shown in Figure \ref{rotacion}, where the
solid lines are fits to Eqs. \ref{ml.mt}. The small discrepancies between
the data and the fits are due to the fact that the ZFC procedure is not
perfect and thus there is always a small remanent field. This field induces
a remanent magnetic moment in the sample that adds to the Meissner signal in
Eqs. \ref{ml.mt}. The periodicity of this remanent moment is $360^{\circ }$,
instead of the $180^{\circ }$ periodicity of the Meissner signal. When that
effect is taken into account, we obtain excellent fits to the data in fig. 
\ref{rotacion}, as described elsewhere \cite{casa}.

A more exact method is to determine the longitudinal and transverse Meissner
slopes $m_l^{\prime }=\frac{dm_l}{dH}$ and $m_t^{\prime }=\frac{dm_t}{dH}$
from the initial part of the magnetization loops ${\bf M}\left( {\bf H}%
\right) $. This procedure eliminates the uncertainty in the value of $H$ due
to the remanent field, and also that arising from remanent magnetic moments,
as they are field independent. In the inset of figure \ref
{rotacion} we have plotted $m_l^{\prime }$ and $m_t^{\prime }$ as a function
of $\Theta _H$ for all the angles measured. We obtain excellent agreement
with eqs. \ref{ml.mt} with the parameters $V=2.29\times 10^{-4}cm^3$ and $%
\nu =0.018$. The volume so obtained coincides within $1\%$ with the volume
calculated by weighting, and it is the value that we have used to calculate
the curves of figure \ref{mlmt} and all the magnetization data shown from
now on. The demagnetizing factor coincides very well with the expected
value $\nu \simeq t/L=0.016$.

It is also useful to calculate the modulus and orientation of ${\bf M}$ in
the Meissner state from Eqs. \ref{ml.mt}. If we call $\alpha $ the angle
formed by ${\bf M}$ and ${\bf H}$, then $\tan \alpha =\left( m_t^{\prime
}/m_l^{\prime }\right) $ is

\begin{equation}
\label{tan.alfa}\tan \alpha =\frac{(1-2\nu )\tan {\Theta _H}}{1+2\nu \tan {%
\Theta _H^2}} 
\end{equation}

while the modulus is

\begin{equation}
\label{mod.Meiss}-4\pi M=\left[ \frac{\sin {\Theta _H^2}}{(1-\nu )^2}+\frac{%
\cos {\Theta _H^2}}{(2\nu )^2}\right] ^{1/2}H 
\end{equation}

We found that Eqs. \ref{ml.mt} and \ref{tan.alfa} allow us to determine the
angle $\Theta _H$ from the initial slopes of our ${\bf M}\left( {\bf H}%
\right) $ data with accuracy better than $1^{\circ }$. As this error is
lower than that associated with our mechanical devise ($\sim 2^{\circ }$),
we decided to use in all cases the value of $\Theta _H$ obtained from the
Meissner slopes, and take the reading of the rotator as a double-check.

\subsection{The reversible response in the mixed state}

\label{sec.rev}In this section we will show that the reversible response of
the sample behaves as expected in an isotropic superconductor.
Assuming that the Bean model is valid, both the longitudinal and transverse
components of the equilibrium or reversible magnetization ${\bf M}_{eq}$ in
the mixed state, $M_{eq,l}$ and $M_{eq,t}$, can be determined by simply
taking the mean value between the two branches of the hysteresis loops, as
illustrated in figure \ref{mlmt}. We see that for fields above $\sim 500Oe$
the
transverse component $M_{eq,t}\approx 0$. This is true for all $\Theta _H$,
i.e., at high fields ${\bf M}_{eq}$ is parallel to ${\bf H}$ regardless of
the sample orientation.

In figure \ref{mmvsh} we show the modulus $M_{eq}$ as a function of $H$ for
several field orientations. Also shown are the corresponding Meissner
responses, which are angle dependent as expected from Eq. \ref{mod.Meiss}.
On the other hand, in the mixed state at high fields (above $\sim 500Oe$)
all the curves coincide; and in particular the condition $M_{eq}\approx 0$
that determines the upper critical field $H_{c2}$ is independent of $\Theta
_H$.

The behavior of ${\bf M}_{eq}$ in the mixed state at high fields is a
consequence of the isotropy of the material. Indeed, the free energy $F$ of
the vortex system in isotropic superconductors is only a function of $B$,
and the equilibrium state is obtained by minimization of the Gibbs free
energy [see Eq. 8.60 in \cite{blatter-et-al-1994}]

\begin{equation}
\label{G}G=F(B)-\frac{B^2}{8\pi }+\frac{\left( {\bf B-H}\right) {\bf M}_{eq}}%
2 
\end{equation}
at high fields $4\pi M_{eq}\ll H$, then ${\bf B\approx H}$, and the last
term in Eq. \ref{G} is negligible. We thus recover the solution of an
infinite superconductor, where ${\bf B}$, ${\bf H}$ and ${\bf M}_{eq}$ are
parallel and $M_{eq}(H)$ is given by the usual expressions\cite{parks}
regardless of the sample geometry.

The penetration of the first vortex becomes energetically favorable when
the modulus of the internal field, $H_i$, equals the lower critical field $%
H_{c1}$. Taking into account that in the Meissner state ${\bf {H}=(\overline{%
\overline{1}}-\overline{\overline{\nu }}){H_i}}$, we can define an
''apparent'' lower critical field $H_{c1}^{*}\left( \Theta _H\right) $ that,
in spite of the material isotropy, is angle dependent due to geometrical
effects:

\begin{equation}
\label{hc1.est}H_{c1}^{*}=H_{c1}\left[ \frac{\sin {\Theta _H}^2}{(1-\nu )^2}+%
\frac{\cos {\Theta _H}^2}{(2\nu )^2}\right] ^{-1/2} 
\end{equation}

The experimental determination of $H_{c1}^{*}\left( \Theta _H\right) $ is
complicated by the fact that for fields only slightly above the Meissner
regime, the critical state is not fully developed in the increasing branch
of the loop (see figure \ref{mlmt}), and the mean magnetization does not give
us a good estimate of $M_{eq}$. To solve this difficulty we adjusted the $%
M_{eq}\left( H\right) $ data at intermediate fields by a dependence
$M_{eq}\propto \ln H$. Although this logarithmic dependence is only expected
in high $\kappa $ superconductors\cite{blatter-et-al-1994,parks}, while in
the present case $\kappa \sim 2.2$, the experimental fact is that the fit
works pretty well. The upper limit of the field range of validity of the fit
is about $H\simeq 500Oe$. The lower limit depends on $\Theta _H$, varying
from $H\simeq 150Oe$ for $\Theta _H\simeq 0^{\circ }$ to $H\simeq 250Oe$ for
$\Theta _H$ approaching $90^{\circ }$. We then
identified the field at which the extrapolation of the fits intersect the
Meissner slope with $H_{c1}^{*}\left( \Theta _H\right) $, as shown in figure 
\ref{mmvsh}.

Evidence that the extrapolations provide good estimates of $%
M_{eq}$ come from the observation that the intersection with the Meissner
slopes occur at approximately the same magnetization value
$4\pi M_{eq}\sim 290Oe$ for all angles. This is what we would expect, as
from Eqs. \ref{mod.Meiss}
and \ref{hc1.est} results $4\pi M_{eq}(H_{c1}^{*})=H_{c1}$, independent of $%
\Theta _H$.

\subsection{The vortex orientation}

We can now use the ${\bf M}_{eq}$ data to calculate the equilibrium
induction field ${\bf B}_{eq}$ according to Eq. \ref{bhm}. In
figure \ref{angb} we
plot the angle $\Theta _B$ between ${\bf B}_{eq}$ and ${\bf n}$, as a
function of $H$ for several orientations $\Theta _H$. At high fields $\Theta
_B\approx \Theta _H$, as already discussed, but as $H$ decreases $\Theta _B$
approaches to zero for all orientations.

This behavior is also understood from thermodynamic considerations. From Eq. 
\ref{G} we can obtain the equilibrium condition by minimization with respect
to ${\bf B}_{eq}$, i.e. $\frac{\partial G}{\partial B_{eq,\perp }}=\frac{%
\partial G}{\partial B_{eq,\parallel }}=0$. Combining with Eq. \ref{bhm}
these conditions result

\begin{eqnarray}
\frac{\partial F}{\partial B_{eq,\perp }}-\frac{B_{eq,\perp }}{4\pi }+\frac{%
\left( B_{eq,\perp }-H_{\perp }\right) }{8\pi \nu }=0 \nonumber \\
\frac{\partial F}{%
\partial B_{eq,\parallel }}-\frac{B_{eq,\parallel }}{4\pi }+\frac{\left(
B_{eq,\parallel }-H_{\parallel }\right) }{4\pi \left( 1-\nu \right) }=0\nonumber \\ 
\end{eqnarray}

For $\nu \ll 1$, the first expression implies that $B_{eq,\perp }\approx
H_{\perp }$ and the second condition reduces to $\frac{\partial F}{\partial
B_{eq,\parallel }}\approx \frac{H_{\parallel }}{4\pi }$. For fields slightly
above $H_{c1}^{*}$, vortex density is very low ($B_{eq}\ll H_{c1}$), then
interactions are negligible and $F\approx \left( B_{eq}/\Phi _0\right)
\varepsilon _l$, where $\varepsilon _l$ is the vortex line energy \cite{blatter-et-al-1994}. In this limit $B_{eq,\parallel }/H_{\parallel }\approx B/H_{c1}$,
thus $B_{eq,\parallel }\ll H_{\parallel }$, i.e., the vortex direction is
very close to ${\bf n}$. At intermediate fields $F\approx \frac{B_{eq}^2}{%
8\pi }+\frac{\Phi _0B_{eq}}{2\left( 4\pi \lambda \right) ^2}\ln \left(
H_{c2}/B_{eq}\right) $ and we obtain

$$
B_{eq,\parallel }\approx H_{\parallel }-\frac{\Phi _0}{8\pi \lambda ^2}\frac{%
B_{eq,\parallel }}B\ln \left( \frac{H_{c2}}{eB_{eq}}\right) 
$$

The physical interpretation of these results is straightforward. The normal
component remains almost unchanged due to flux conservation (a result that
of course becomes exact in the limit of an infinite slab). At low fields the
parallel component is reduced because the system gains energy by shortening
the vortex length. As field increases interactions also favor the reduction
of $B_{eq,\parallel }$, but eventually the energy cost of having ${\bf B}%
_{eq}$ non-parallel to ${\bf H}$ becomes too large. Thus, vortex orientation
shifts from ${\bf n}$ to ${\bf H}$ as field increases. The modulus $B_{eq}$
evolves from $B_{eq}\sim H_{\perp }$ at low fields to $B_{eq}\sim H$ at high
fields.

\subsection{The irreversible response}

\label{sec.irrev}According to the Bean model\cite{bean-1962},
the irreversible magnetization
is related to $J_c$ through a geometry dependent proportionality constant.
We define the irreversible magnetization vector as ${\bf M}_{irr}=\frac 12%
{\bf \Delta M}$, where the components of ${\bf \Delta M}$ are the widths of
the longitudinal and transverse hysteresis loops, $\Delta M_l$ and $\Delta
M_t$ respectively. In thin samples the non-equilibrium currents are
constrained to the sample plane, thus they should generate an irreversible
magnetization ${\bf M}_{irr}$ that is almost locked to the sample normal\cite{zhukov-et-al-1997}
in a wide angular regime $0\leq \Theta _H\leq \Theta _c$, where $\tan \left(
\Theta _c\right) \sim L/t$. When $\Theta _H\sim \Theta _c$, the direction of 
${\bf M}_{irr}$ is expected to rotate $\sim 90^{\circ }$ in a narrow angular
range and to locate in the plane of the plate. This behavior has been
recently demonstrated by Zhukov et al. \cite{zhukov-et-al-1997} in detailed
studies of the modulus and orientation of the irreversible magnetization of
both isotropic and anisotropic thin superconductors.

We have confirmed that in our samples, for $\Theta _H\leq 70^{\circ }$ and
for all applied fields, ${\bf M}_{irr}$ points in the direction of ${\bf n}$
within our $1^{\circ }$ resolution. For angles $\Theta _H\geq 70^{\circ }$
the situation is more complex. ${\bf M}_{irr}$ progressively tilts away from 
${\bf n}$ as $\Theta _H$ increases, but its direction is field dependent.
This behavior cannot be described by the expressions proposed by Zhukov et
al.\cite{zhukov-et-al-1997}, where the field dependence of $J_c$ is not
considered.

In figure \ref{dmvsh}(a) we show the modulus $M_{irr}$ as a function of $H$ for
different angles $\Theta _H$. For clarity we only show data up to $\Theta _H$
slightly above $70^{\circ }$, thus all the curves in the figure correspond
to ${\bf M}_{irr}\parallel {\bf n}$ or very close to it. We first note that $%
M_{irr}$ goes to zero for all $\Theta _H$ at $H=H_{c2}$, as expected for an
isotropic type-II superconductor. Another evident result is that, although
our superconductor is isotropic, $M_{irr}$ is angle dependent, even for $%
\Theta _H\leq 70^{\circ }$. We will now show that this dependence originates
in sample geometry effects.

Clearly, to understand these results we must go beyond the Bean model and
consider the field dependence of $J_c$. There are two basic points to be
taken into account. The first one is that $J_c$ (and consequently $M_{irr}$)
is a function of $B$ rather than $H$. The second one is that the
non-equilibrium currents determine the direction of the {\it variation} of $%
{\bf B}\left( {\bf r}\right) $ (through ${\bf J\propto \nabla \times B}$)
but not the vortex orientation itself. Indeed, the field distribution in the
critical state of thin samples, either for $\Theta _H=0$ or for $\Theta
_H\neq 0$, has been calculated taking into account the contribution of the
irreversible currents plus the applied field, but not the reversible
magnetization\cite{clem-1994}.

We will consider that the field ${\bf B}\left( {\bf r}\right) $ arises from
the superposition of the spatially inhomogeneous contribution produced by
the non-equilibrium currents and the homogeneous reversible contribution
calculated in the previous section. Provided that the irreversible
contribution to ${\bf B}\left( {\bf r}\right) $ is not too large as compared
to ${\bf B}_{eq}$, as indeed occurs in our samples, then the average vortex
orientation is still approximately given by ${\bf B}_{eq}$. This also allows
us to make the approximation that $M_{irr}$ is simply a function of $B_{eq}$%
. (This is equivalent to the usual approach of considering that, although $%
J_c$ is field dependent, it is almost constant over the range of field
variation within the sample for a given $H$, and thus the Bean model can be
applied to each field). At low fields we can further
approximate $B_{eq}\sim $ $H_{\perp }$, thus we expect $M_{irr}$ to be a
function of $H_{\perp }$.

To check the above modeling, in figure \ref{dmvsh}(b) we re-plot the 
$M_{irr}$ data of fig. \ref{dmvsh}(a), as a function of $H_{\perp }$. The good
scaling of the curves in the low field range indicates that the angular
dependence of $M_{irr}$ in this isotropic superconductor can be understood
in terms of the angular-dependent equilibrium vortex orientation.

In figure \ref{dmvsang} we show $M_{irr}$ as a function of $\Theta _H$ for
several $H$. At very low fields ($H\leq 10Oe$), $M_{irr}$ is almost constant
over a wide angular range, starts to decrease at around $\Theta _H\sim
70^{\circ }$ and reaches its minimum value at $\Theta _H=90^{\circ }$. This
is, at least qualitatively, the expected behavior\cite{zhukov-et-al-1997} in
thin samples when the field dependence of $J_c$ is not taken into account.
Indeed, the dotted line is the dependence predicted by Zhukov et al
\cite{zhukov-et-al-1997}, where we have used the aspect ratio $\nu \sim 0.018$ of
our sample determined previously, and thus there are no free parameters. We
note, however, that a better fit could be obtained using a larger $\nu $.

At higher fields we must take into account the vortex orientation as
described above. On the other hand, for $\Theta _H\leq 70^{\circ }$ we do
not need to consider the reduction of $M_irr$ due to the geometrical effects
described in \cite{zhukov-et-al-1997}. We thus expect an angular dependence

\begin{equation}
\label{ec.scaling}M_{irr}(H,\Theta _H)=M_{irr}(H_{\perp },\Theta
_H=0)=M_{irr}^0(H\cos {\Theta _H})
\end{equation}

where $M_{irr}^0$ is the curve measured for $\Theta _H=0$. The validity of
this scaling is manifested in the figure \ref{dmvsang}. An excellent agreement
between the experimental data and eq. \ref{ec.scaling} is observed for  $%
\Theta _H\leq 70^{\circ }$ in the field range $0\leq H\leq 240Oe$, where $B$
is approximately aligned with the normal to the sample. For $\Theta
_H>70^{\circ }$ the decrease of $M_{irr}$ due to sample geometry takes over.
For fields above $240Oe$, the fit fails because the $B_{eq}\sim $ $H_{\perp }
$ approximation is not satisfied. We note that this model satisfactorily
accounts for the angular dependence of $M_{irr}$ that Zhukov et
al.\cite{zhukov-et-al-1997} observed at high fields in their isotropic
samples, which could not be explained within their picture.

\section{Conclusion}

We have shown that the variable that governs the angle dependence of $M_{irr}
$ in isotropic thin samples is the magnetic induction ${\bf B}$, which
describes the density and orientation of the vortices. The behavior of $%
M_{irr}\left( H,\Theta _H\right) $ at low fields and over a wide angular
range can be understood by simultaneously taking into account the
modulus and orientation of the equilibrium magnetization vector ${\bf B}_{eq}
$, and the field dependence of $M_{irr}$ when ${\bf H}$ is normal to the
sample. To perform the analysis reported here, it was necessary to
simultaneously measure both components of the magnetization in an isotropic
sample of well defined geometry, and to be able to decompose it in the
reversible and irreversible parts. This last step was only possible because
we used a sample with very low $J_c$.

It is a well known experimental fact \cite{roas90} that HTSCs exhibit the same
scaling of $M_{irr}$ with $H_{\perp }$ as described by Eq. \ref{ec.scaling}.
That behavior is consistent with the anisotropic scaling expected for these
materials in the limit of very large anisotropy\cite{blatter-et-al-1992,blatter-et-al-1994},
and is usually taken as evidence for quasi two dimensional behavior.
However, we have observed that the same scaling occurs in our isotropic
sample, thus indicating that geometrical effects should be carefully taken
into account in those studies.

\begin{figure}[htb]
\caption[]{(a) Open symbols: Transverse component of the magnetization as a
function of applied field for the sample normal at an angle of
$\sim 60^{\circ}$. The arrows indicate the direction of the field sweep.
Full symbols: Mean value of the magnetization. (b) Idem for the longitudinal
component. The sketch shows the components and angles used in the text.}
\label{mlmt} 
\end{figure}

\begin{figure}
\caption[]{Angular dependence of the longitudinal and transverse components
of the magnetization in the Meissner state, as measured by rotating the
sample at fixed applied field (the origin of angles is arbitrary). The solid lines are fits to Eq. \ref{ml.mt}.
Inset: Meissner slopes as a function of angle. The solid lines are fits to
Eq. \ref{ml.mt}, see text.}
\label{rotacion} 
\end{figure}

\begin{figure}[htb]
\caption[]{Modulus of the equilibrium magnetization in the Meissner and mixed
states as a function of applied field, for several sample orientations. The
dotted lines are fits $M_{eq}\propto \ln H$. The inset shows a blow up of the
same data in semilogarithmic scale. The intersections of the extrapolations
of the logarithmic fits with the Meissner response (shown by crosses) provide
an estimate of $H_{c1}^{*}\left( \Theta _H\right) $}
\label{mmvsh} 
\end{figure}

\begin{figure}
\caption[]{Angle between the equilibrium induction field ${\bf B}$ and the
normal to the sample, as a function of applied field, for three sample
orientations $\Theta _H$. The symbols are experimental data; the dotted
lines are the extrapolations of the logarithmic fits shown in figure 3.}
\label{angb} 
\end{figure}

\begin{figure}
\caption[]{(a): Irreversible magnetization as a function of applied field for several sample orientations. (b): Same data, as a function of the normal
component of the applied field.}
\label{dmvsh}
\end{figure}

\begin{figure}
\caption[]{Irreversible magnetization as a function of $\Theta _H$ for
several applied fields. From top to bottom, $H=0, 20, 30, 40, 50, 60, 70, 80, 100, 120, 160, 200, 240Oe$. The solid lines are fits to Eq. \ref{ec.scaling}. The dotted line is a fit to the model of Ref. \cite{zhukov-et-al-1997} using an aspect ratio $\nu = 0.018$}
\label{dmvsang} 
\end{figure}


\begin{references}

\bibitem{clem-1994} Clem J R and Sanchez A 1994 {\it Phys.\ Rev.\ B.} {\bf 50} 9355

\bibitem{brandt-1994b} Brandt E  H  1994  {\it Phys.\ Rev.\ B.} {\bf 49}  9024; 1994 {\it Phys.\ Rev.\ B.} {\bf 50}  4034; 1995 {\it Phys.\ Rev.\ B.} {\bf 52}  15442  

\bibitem{schuster93a}  Schuster T, Kuhn H, Koblischka M  R, Theuss H,  Kronm{\"u}ller H, Leghissa M, Kraus M  and Saemann-Ischenko G 1993 {\it Phys.\ Rev.\ B.} {\bf 47}  373; Schuster Th, Kuhn H, Brandt E  H, Indenbom M  V, Kl{\"a}ser M, M{\"u}ller-Vogt G, Habermeier H-U, Kronm{\"u}ller H   and Forkl A   1995 {\it Phys.\ Rev.\ B.} {\bf 52}  10375; Schuster T, Leghissa M, Koblischka M  R, Kuhn H, Kraus M, Kronm{\"u}ller H  and  Saemann-Ischenko G   1992 {\it Physica C}  {\bf 203}  203  

\bibitem{frankel-1979} Frankel D  J  1979 {\it J.\ Appl.\ Phys.} {\bf 50}  5402   

\bibitem{kunchur-et-al-1991} Kunchur  M  N  and Poon  S  J  1991 {\it Phys.\ Rev.\ B.} {\bf 43}  2916   

\bibitem{hellman-et-al-1992} Hellman F,  Gyorgy  E  M  and  Dynes R  C  1992  {\it Phys.\ Rev.\ Lett.} {\bf 68}  867  

\bibitem{zhukov-et-al-1997} Zhukov A  A, Perkins G  K, Bugoslavsky  Yu  V  and  Caplin A  D  1997  {\it Phys.\ Rev.\ B.} {\bf 56}  2809   

\bibitem{parks} Parks R  D  (ed.) 1969  {\it Superconductivity} Marcel Dekker  Inc., New York   

\bibitem{landau-v8} Landau  L  D  and  Lifshitz E  M   1960 {\it Electrodynamics of Continuous Media} Pergamon Press Ltd., Oxford   

\bibitem{campbell-1972} Campbell A  M  and  Evetts J  E  1972  {\it Adv.\ Phys.} {\bf 21}  199   

\bibitem{bean-1962}  Bean C  P  1962  {\it Phys.\ Rev.\ Lett.} {\bf 8}  250  

\bibitem{daeumling-1989} Daeumling M  and  Larbalestier D  C  1989 {\it Phys.\ Rev.\ B.} {\bf 40}  9350

\bibitem{conner-1991} Conner L  W  and  Malozemoff A  P  1991 {\it Phys.\ Rev.\ B.} {\bf 43}  402

\bibitem{tamegai-1992} Tamegai T, Krusin-Elbaum L, Santhanam P, Brady M  J, Feild C  and Holtzberg F  1992 {\it Phys.\ Rev.\ B.} {\bf 45}  2589

\bibitem{blatter-et-al-1992} Blatter G, Geshkenbein V  B  and Larkin A  I  1992 {\it Phys.\ Rev.\ B.} {\bf 6}  875

\bibitem{blatter-et-al-1994} Blatter G, Feigel'man M  V, Geshkenbein V  B, Larkin A  I  and Vinokur V  M  1994  {\it Rev.\ Mod.\ Phys.} {\bf 4}  1125  

\bibitem{buzdin-et-al-1990} Buzdin  A  I  and  Simonov A  Yu  1991  {\it Physica C} {\bf 175}  143

\bibitem{casa} Casa D  M  et al.  {\it to be published}

\bibitem{qd} {\it Quantum Design operating manual and application notes} 

\bibitem{roas90} Roas B, Schultz L and Saemann-Ischenko G, 1990 {\it Phys.\ Rev.\ Lett.} {\bf 64} 479; Schimtt P, Kummeth P, Schulz L and Saemann-Ischenko G 1991 {\it Phys.\ Rev.\ Lett.} {\bf 67} 267 

\end{references}
\end{document}